\documentclass[preprint,12pt,showkeys,nofootinbib,floatfix]{revtex4}
\usepackage{amssymb}
\usepackage{amsfonts}
\usepackage{amsmath}
\usepackage{graphicx}
\usepackage{subfigure}
\usepackage[dvipdfm,
            pdfstartview=FitH,
            bookmarksnumbered=true,
            bookmarksopen=true,
            colorlinks,
            pdfborder=001,
            linkcolor=green,
            anchorcolor=green,
            citecolor=red
            ]{hyperref}
\usepackage[toc,page,title,titletoc,header]{appendix}
\usepackage{graphics}
\usepackage{epsfig}
\usepackage{slashed}
\usepackage{latexsym}
\usepackage{syntonly}
\begin{document}
\title{A holographic model of $d$-wave superconductor vortices with Lifshitz scaling}
\author{Hong Guo$^{1}$}
\author{Fu-Wen Shu$^{1,2}$}
\thanks{E-mail address: shufuwen@ncu.edu.cn}
\author{Jing-He Chen$^{1}$}
\author{Hui Li$^{1}$}
\author{Ze Yu$^{1}$}
\affiliation{
$^{1}$Department of Physics, Nanchang University, Nanchang, 330031, China\\
$^{2}$Center for Relativistic Astrophysics and High Energy Physics, Nanchang University, Nanchang, 330031, China}
\begin{abstract}
 We study analytically the $d$-wave holographic superconductors with Lifshitz scaling in the presence of external magnetic field. The vortex lattice solutions of the model have also been obtained with different Lifshitz scaling. Our results imply that holographic $d$-wave superconductor is indeed a type II one even for different Lifshitz scaling. This is the same as the conventional $d$-wave superconductors in the Ginzburg-Landau theory.
  Our results also indicate that the dynamical exponent $z$ has no effect to the shape of the vortex lattice even after higher order corrections (away from the phase transition point $B_c$) are included. However, it has effects on the upper critical magnetic field $B_{c_2}$ through the fact that a larger $z$ results in a smaller $B_{c_2}$ and therefore influences the size (characterized by $r_0\equiv 1/\sqrt{B_{c_2}}$) of the vortex lattices. Furthermore, close comparisons between our results and those of the Ginzburg-Landau theory reveal the fact that the upper critical magnetic field $B_{c_2}$ is inversely proportional to the square of the superconducting coherence length $\xi$, regardless of the anisotropy between space and time.
\end{abstract}

\maketitle
%%\maketitle  IS IGNORED %%%%%%%%%%%
\section{Introduction}
The gauge-gravity duality  \cite{ads/cft,gkp,w} offers a very
promising way to explore the possible dynamics of strongly
interacting matter in field theory. It provides a well-established
method for calculating correlation functions in a strongly
interacting field theory in terms of a dual classical gravity
description. While its relevance to any specific strong coupling
system that can realized in the laboratory is not well-understood,
it nonetheless provides a window through which we might hope to
obtain insight into the properties of some condensed matter systems
that defy description by traditional approaches. One of the unsolved
mysteries in modern condensed matter physics is the mechanism of the
high temperature superconductors cuprates(HTSC). These materials are
layered compounds with copper-oxygen planes and are doped Mott
insulators with strong electronic correlations which the pairing
symmetry is unconventional and there is a strong experimental
evidence showing that it is the $d$-wave superconductor. This makes
the $d$-wave superconductor particularly attractive for physicists.
%It is speculated that the pairing between electrons is mediated via strong antiferromagnetic spin fluctuations in the system. A promment strong-coupling theory called the resonant valence bond(RVB) theory describing liquid state with spin singlets is proposed by Anderson. Upon hole doping, the N\'{e}el order is destroyed and give rise to superconductivity\cite{Anderson}. Several gauge theories have been proposed to formulate the RVB physics, by enforcing the double occupation constraint in the strong coupling limit\cite{gaugetheories}.

It was Gubser who first noticed that by coupling the Abelian Higgs model to gravity with a negative cosmological constant, one can find solutions that spontaneously break the Abelian gauge symmetry via a charged complex scalar condensate near the horizon of the black hole\cite{gub1,gub2}. This model exhibits the key properties of superconductivity: a phase transition at a critical temperature, where a spontaneous symmetry
breaking of a $U(1)$ gauge symmetry in the bulk gravitational theory corresponds to a broken global $U(1)$ symmetry on the boundary, and the formation of a charged condensate. Based on this observation, Hartnoll et al proposed a holographic model for $s$-wave superconductors by considering a neutral black hole with a charged scalar and the Maxwell field\cite{horowitz}. Since then this correspondence has been widely explored in order to understand several crucial properties of these
holographic superconductors (see Ref. \cite{HH} for reviews). The gravitational model that dual to the $d$-wave superconductors was proposed in \cite{wen,be} where the complex scalar field  for the $s$-wave model is replaced by a symmetric traceless tensor.

One of the major characteristic properties of superconductors is that they expel magnetic fields as the temperature is
lowered through the critical temperature. In the presence of an external magnetic field, ordinary superconductors may
be classified into two categories, namely type I and type II. It was found that at $T<T_c$, magnetic field expels the wave condensation for holographic $s$-wave superconductor\cite{mps,Maeda2}, holographic $p$-wave superconductor \cite{murray}, and holographic $d$-wave superconductor\cite{arXiv:1006.5483} as well, along with the formation of Abrikosov vortices.
This indicates that these holographic models of superconductor belong to type II ones. However, all these holographic models were constructed only in the relativistic spacetimes. Thus
we wonder whether the holographic $d$-wave superconductor still be the type II one in non-relativistic spacetimes, for example, Lifshitz spacetime, which is our main motivation in this paper.

It is often observed that the behaviors of many condensed matter systems are governed by Lifshitz-like fixed points. These fixed points are characterized by the anisotropic scaling symmetry
$$
t\rightarrow \lambda^zt,\ \ \ x^i\rightarrow \lambda x^i,
$$
where $z$ is called the ¡°dynamical critical exponent¡± and it describes the degree of anisotropy between space and time. The nonrelativistic nature of these systems makes the dual description different and a gravity dual for such systems can be realized by nonrelativistic CFTs \cite{dtson,kbm,wdg,taylor}. Recently, Bu used the nonrelativistic AdS/CFT correspondence to study the holographic superconductors in the Lifshitz black hole geometry for $z=2$ in order to explore the effects of the dynamical exponent and distinguish some universal properties of holographic superconductors \cite{Bu}. It is found that the Lifshitz black hole geometry results in different asymptotic behaviors of temporal and spatial components of gauge fields compared to those in the Schwarzschild-AdS black hole, which brings some new features of holographic superconductor models. More recently, Lu $\emph{et al}.$ discussed the effects of the Lifshitz dynamical exponent $z$ on holographic superconductors and gave some different results from the
Schwarzschild-AdS background \cite{Lu}. To date, there have attracted considerable interest to generalize the holographic superconducting models to nonrelativistic situations \cite{CaiLF,JingLF,Sin,Brynjolfsson,MomeniLifshitz,Schaposnik,Abdalla,Tallarita}.

In this paper,we analytically study the spatially dependent equations of motion for the $d$-wave holographic superconductor with Lifshitz scaling when the added magnetic field is slightly below the upper critical magnetic field. We want to distinguish the effects of the dynamical exponent to the vortex lattice and explore the behavior of the upper critical magnetic field. In particular, according to the Ginzburg-Landau (GL) theory, it should be noted that the upper critical magnetic field has the well-known relation $B_{c}\propto(1-T/T_c)$ \cite{Poole}. A number of attempts have been made to investigate the effects of applying an external magnetic field to holographic dual models \cite{SetareEPL,GeBackreac,Montull,Gao,Ged,Roychowdhury,Momeni,Roychowdhury2,Cui,RoychowdhuryJHEP2013,Cai2013,Gangopadhyay,Lala}. All these papers are made in relativistic situations. It is therefore very natural to consider the nonrelativistic situations, such as Lifshitz black hole. Furthermore, we constructed the vortex lattice solution, or the Abrikosov lattice which is characterized by two lattice parameters, $a_1$ and $a_2$, perturbatively near the second-order phase transition. There is an observation\cite{Lala} that the dynamical exponent has no effect to the shape of the vortex for $s-$wave superconductor. In this paper, one of main motivation is to see whether it is still correct for the $d-$wave superconductor. In addition, we would pay much attention to see this point a little bit away from the second-order phase transition point. Our result is very interesting, it seems that the exponent $z$ does not influence the shape of the lattice for higher order corrections.

The organization of this paper is as follows. In section 2, we will study the $d$-wave holographic superconductors with Lifshitz scaling. In section 3 we investigate the properties of the holographic superconductors with Lifshitz scaling in an external magnetic field. Section 4 is devoted to the construction of vortex solution of the $d$-wave model and to show that the dynamical exponent does not influence the shape of the vortex. And we will conclude in the last section of our main results.

\section{The $d$-wave holographic superconductor models with Lifshitz scaling}
In this section we first give the spatial dependent equations of motion for the $d$-wave model in the presence of a uniform magnetic field, then we will study the condensate solution and discuss the critical temperature.

\subsection{Holographic $d$-wave superconductor: the model}
The action of the $d$-wave superconductor in $4$ dimensions is the following\footnote{In principle, it is possible to generalize our analysis to higher dimensions.}\cite{wen}
\begin{eqnarray}
&&S=\frac{1}{2\kappa^2}\int d^4x \sqrt{-g}\bigg\{(R-2\Lambda)+\mathcal{L}_m\bigg\},\\
&&\mathcal{L}_m=-\frac{L^2}{q^2}\bigg[(D_{\mu}B_{\nu\gamma})^{*}D^{\mu}B^{\nu\gamma}+m^2B_{\mu\nu}^{*}B^{\mu\nu}+\frac{1}{4}F_{\mu\nu}F^{\mu\nu}\bigg]
\end{eqnarray}
where $B_{\mu\nu}$ is a symmetric traceless tensor, $R$ is the Ricci scalar, $\Lambda=-\frac{3}{L^2}$ is the negative cosmological constant with $L$ the AdS radius, and $\kappa^2=8\pi G_N$ is the gravitational coupling. $D_{\mu}=\partial_{\mu}+iA_{\mu}$ is the covariant derivative, $q$ and $m^2$ are the charge and mass squared of $B_{\mu\nu}$, respectively.

Working in the probe limit in which the matter fields do not backreact on the metric and taking the planar Lifshitz-AdS ansatz, the black hole metric reads:
\begin{equation}
ds^2=L^2\bigg(-r^{2z}h(r)dt^2+\frac{1}{r^2h(r)}dr^2+r^2dx^2+r^2dy^2\bigg)
\end{equation}
where the metric coefficient
\begin{equation}
h(r)=1-\frac{r^{z+2}_{+}}{r^{z+2}},
\end{equation}
and $r_{+}$ is the horizon radius of the black hole. The
Hawking temperature of the black hole is $T=\frac{(z+2)r^z_{+}}{4\pi L^2}$. Setting $u=\frac{r_{+}}{r}$, the metric can be rewritten
in the form
\begin{equation}
ds^2=L^2\bigg(-\frac{r^{2z}_{+}}{u^{2z}}h(u)dt^2+\frac{du^2}{u^2h(u)}+\frac{r^2_{+}}{u^2}dx^2+\frac{r^2_{+}}{u^2}dy^2\bigg),
\end{equation}
where $h(z)=1-u^{z+2}$.

The $d$-wave superconductors is translational invariant on the $(x-y)$ plane and condensate on the boundary, while the rotational symmetry is broken down to $Z(2)$ due to the condensate change its sign under a $\pi/2$ rotation on the $(x-y)$ plane. To fulfill these features we take the following ansatz\cite{wen}
\begin{eqnarray}
B_{\mu \nu }=\text{diagonal}\left( 0,0,f(u,x,y),-f(u,x,y)\right) ,~~A=\phi(u,x,y)dt+A_y(u,x,y)dy,
\end{eqnarray}
where we keep a nonvanishing $A_y$ so as to have an external magnetic field along $y$ direction.

After variation of the action with this ansatz, the equations of
motion for the tensor field $B_{\mu\nu}$, the gauge field components
$A_t$ and $A_y$ are given, respectively, by
\begin{subequations}
\begin{eqnarray}\label{eom1}
&&h\partial_{u}^2 f + (\partial_{u}h + \frac{3-z}{u}h)\partial_{u} f + \frac{1}{r^2_{+}}(\partial_{x}^2 f + \partial_{y}^2 f) + \frac{2iA_y}{r^2_{+}}\partial_{y} f + \frac{i \partial_{y} {A_y}}{r^2_{+}}f + \frac{2\partial_{u} h}{u}f \nonumber \\  &&+ \frac{u^{2z-2} \phi^2}{r^{2z}_{+} h}f- \frac{2(z+1)h}{u^2}f-\frac{A^2_y}{r^2_{+}}f-\frac{L^2m^2}{u^2}f=0,\\
&&h\partial_{u}^2 \phi+\frac{1}{r^2_{+}}(\partial_x^2+\partial_y^2)\phi+\frac{(z-1)h}{u}\partial_u \phi-\frac{4u^2 \mid f\mid ^2\phi}{r^4_{+}L^2}=0,\\
\label{eom2}&&h\partial_u^2 A_y+(\partial_u h-\frac{z-1}{u}h)\partial_u A_y+\frac{1}{r^2_{+}}\partial_x^2 A_y+\frac{2iu^2f^*\partial_y f}{r^4_{+}L^2}-\frac{2iu^2f\partial_y f^*}{r^4_{+}L^2}-\frac{4u^2A_y \mid f\mid ^2}{r^4_{+}L^2}=0.\nonumber\\
\end{eqnarray}
\end{subequations}
\subsection{Condensate in holographic $d$-wave superconductors without external magnetic filed}
The equations of motion (\ref{eom1})-(\ref{eom2}) of the $d$-wave superconductors are very similar to the $s$-wave model and the matching method should be valid. In order to solve the above equations, let us impose the boundary condition near the horizon and in the asymptotic AdS region, respectively:\\
1). On the horizon $u=1$, as usual, we must have $\phi=0$ so that $\phi dt$ is well defined and the other fields are regular.\\
2). At infinity $u=0$, the solution of fields behaves like
\begin{subequations}
\begin{eqnarray}
 f(u) &=& J_{-}u^{\Delta_{-}}+J_{+}u^{\Delta_{+}},\label{bd1}\\
 \phi(u)& =& \mu-\rho\frac{u^{2-z}}{r^{2-z}_{+}}+\cdots,\label{bd2}\\
 B(\textbf{x}) &=& \partial_x A_y-\partial_y A_x,
\end{eqnarray}
\end{subequations}
where $\Delta_{\pm}=\frac{-(2-z)\pm\sqrt{(2-z)^2+8(z+1)+4m^2L^2}}{2}$. The coefficients $J_{-}$ represents as the source of the dual operator and $J_{+}$ correspond to the vacuum expectation values of the operator that couples to $B_{\mu\nu}$ at the boundary theory. BF bound requires $m^2L^2\geq -\frac{(2-z)^2+8(z+1)}{4}$ (thus $\Delta_{+}\geq 0$) such that the $J_{+}$ term is a constant or vanishes on the boundary.

To solve the critical temperature with the spatial dependent equations of motion, we  ignore the influence of the external magnetic field so as to get the equations of motion only with the reaction of radial coordinates:
\begin{subequations}
\begin{eqnarray}
\label{a1}f''+\left(\frac{h'}{h}+\frac{3-z}{u}\right)f'+\left(\frac{2h'}{uh}+\frac{u^{2z-2}\phi^2}{r_{+}^{2z}h^2}-\frac{2(z+1)}{u^2}-\frac{m^2L^2}{u^2h}\right)f&=&0, \\
  \phi''+\frac{z-1}{u}\phi'-\frac{4 u^2 \left|f\right|^2}{r^{4}_{+}L^2 h}\phi &=& 0 \label{a2} .
\end{eqnarray}
\end{subequations}
As we can see, the change of the equations does not affect the
boundary conditions. We impose boundary condition $J_{-}=0$ in the
following discussion. For  clarify, we set $J=J_{+}$ and
$\Delta=\Delta_{+}$ in this work.

It should be noted that Frobenius analysis of the equations of
motion near the boundary reveals that $\phi(u)=\mu-\rho\log{u}$ for
the case $z=d$. For simplicity, we will not consider this case in
the following studies.

Following the matching method applied in \cite{Pan}, which expands the fields $f$ and $\phi$ near the horizon $u=1$, reads off the expanded solutions from the equations of motion with the above boundary conditions, then matchs the asymptotic solutions at some intermediate point $u=u_m$, in the end we obtain
\begin{equation}
 J=\frac{u^{1-\Delta}_m\left[2(z+2)(2-u_m)+m^2L^2(1-u_m)\right]}{(z+2)\left[(1-u_m)\Delta+2u_m\right]}f(1),
\end{equation}
where
\begin{equation}
 f^2(1)=\frac{(z+2)\left[1+(1-z)(1-u_m)\right]}{4(1-u_m)}\left(\frac{4\pi T}{z+2}\right)^{\frac{4}{z}}\left(\frac{T_c}{T}\right)^{\frac{2}{z}}\left[1-\left(\frac{T}{T_c}\right)^{\frac{2}{z}}\right],
\end{equation}
and we have defined the critical temperature $T_c$
\begin{equation}
 T_c=\frac{z+2}{4\pi}\bigg[\frac{(2-z)\rho u^{1-z}_m}{\alpha\left[1+(1-z)(1-u_m)\right]}\bigg]^{\frac{z}{2}}.\label{T_c}
\end{equation}
The parameter $\alpha$ in \eqref{T_c} is given by
\begin{eqnarray}\label{alpha}
\alpha^2&=&14(z+2)^2+10(z+2)m^2L^2+m^4L^4\nonumber \\
&&+\frac{\left[8(z+2)^2+4(z+2)m^2L^2\right]\left[u_m+(1-u_m)\Delta\right]+4(z+2)^2\Delta}{(1-u_m)\left[2u_m+(1-u_m)\Delta\right]}.
\end{eqnarray}
In order to avoid a breakdown of the matching method, we take the value of $m^2$ to ensure that $\alpha$ is real and find the range of the matching point
\begin{equation}
0<u_m<1, \text{for} -(7-\sqrt{23})(z+2)<m^2<0,
\end{equation}
It is interesting to observe that the value of $-(7-\sqrt{23})(z+2)$ is smaller than $-\frac{(2-z)^2+8(z+1)}{4}$ when $z<2$, which means that the value of $\alpha$ is real all the while when we set the range of $m^2$
\begin{equation}
-\frac{1}{4}\left[(z-2)^2+8(z+1)\right]<m^2<0,
\end{equation}
so it is convenient to use the range $0<u_m<1$ in this work.
In addition, Eq. \eqref{T_c} implies that the larger dynamical exponent $z$ makes the condensation harder to form.

\begin{figure}[htbp]
 \begin{minipage}{1\hsize}
\begin{center}
%\vspace*{10mm}
\includegraphics[scale=0.8]{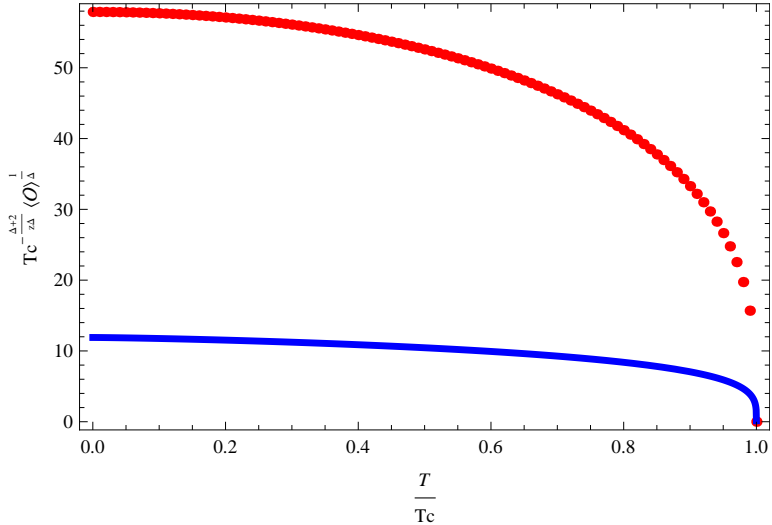}
\end{center}
\caption{ \label{Fig01} The condensation operator as a function of $T/T_c$ obtained by using the analytic matching method. We choose $m^2L^2=-1/4$ and $u_m=1/2$. The top dotted line corresponds to $z=1$ and the bottom one is $z=3/2$.}
\end{minipage}
\end{figure}

According to the AdS/CFT dictionary, near the critical temperature $T\sim T_c$ we can express the relation for the condensation operator $\langle\mathcal{O}\rangle=Jr^{\Delta}_{+}$ as
\begin{eqnarray}
\langle \mathcal{O}\rangle =\bigg(\frac{4\pi
T_c}{z+2}\bigg)^{\frac{2+\Delta}{z}}
\left\{\frac{u^{1-\Delta}_m\left[2(z+2)(2-u_m)+m^2L^2(1-u_m)\right]}{(1-u_m)\Delta+2u_m}\right\}\nonumber \\
\cdot\bigg[\frac{1+(z-1)(1-u_m)}{4(1-u_m)(z+2)}\bigg]^{\frac{1}{2}}\bigg[1-\left(\frac{T}{T_c}\right)^{\frac{2}{z}}\bigg]^{\frac{1}{2}}.
\end{eqnarray}
The analytic result shows that the phase transition of
holographic superconductors with Lifshitz scaling belongs to the
second order. It also indicates that condensation versus temperature have a square root behavior near $T_c$, which suggests
that the critical exponent is 1/2, as expected from the mean field theory. The Lifshitz scaling and spacetime dimension
will not influence the result\cite{Lu}. %It implies that the matching method
%is still powerful to study the $D$-waves holographic superconductors
%in Lifshitz black hole.

In Fig.\ref{Fig01}, we visualize the condensate of the operator $\langle \mathcal{O} \rangle$ as a function of temperature with different dynamical exponent $z$ for the mass of the traceless tensor field $m^2L^2=-1/4$. It is observed that corresponding to the lower critical temperature, the gap becomes increasingly smaller as $z$ increases than the results in \cite{Pan}.

\section{Effects of external magnetic filed on the holographic $d$-wave superconductor}

In this section we would like to study the effect of external magnetic field on the holographic superconductors with Lifshitz scaling. From the gauge/gravity correspondence, the asymptotic value of the magnetic field corresponds to a magnetic field added to the boundary field theory. Near the upper critical magnetic field $B_{c_2}$, the tensor field $f$ can be regarded as a perturbation.
\subsection{Perturbative expansion of the equations of motion}
It is very difficult to exactly solve the above nonlinear coupled partial differential. However, as the magnetic field is slightly below the upper
critical field $B_{c_2}$, it is possible to solve them perturbatively, as what have done in\cite{Maeda2}. For this purpose, we introduce a small
parameter $\epsilon=\frac{B_{c_2}-B}{B_{c_2}}$, and the fields can be then expanded as follows
\begin{subequations}\label{order}
\label{expansion}
\begin{eqnarray}
  &f(\textbf{x},u)&=\epsilon^{1/2}f_1(\textbf{x},u) + \epsilon^{3/2}f_2(\text{x},u) + \cdots,\\
  &A_y(\textbf{x},u)&= A^{(0)}_y(\textbf{x},u) + \epsilon A^{(1)}_y(\textbf{x},u) + \cdots,\\
  &\phi(\textbf{x},u)&=\phi^{(0)}(\textbf{x},u)+\epsilon\phi^{(1)}(\text{x},u)+ \cdots~
\end{eqnarray}
\end{subequations}
where $\textbf{x}=(x,y)$. The zeroth order solution is
\begin{equation}
  f^{(0)}=0,~~\phi^{(0)} = \mu -\rho \left(\frac{u}{r_{+}}\right)^{2-z},~~A^{0}_y = B_{c_2} x,
\label{e4}
\end{equation}
where the rotational symmetry keeps unbroken and therefore it corresponds to the normal state. The magnetic field on the boundary is given by $B=\partial_x A_y-\partial_y A_x=B_{c_2}$ as expected.

Without loss of generality, we assume $f_1(\textbf{x},u)=e^{ipy}\rho_n(u)\gamma_n(x;p)/L^2$ (with $p$ and $\lambda_n$ some constants), then the
equation of motion for $F$ reduces to
%\begin{align}
%  \left[h\partial_{u}^2 +\left(\partial_{u}h+\frac{(3-z)h}{u}\right)\partial_{u}+\frac{2\partial_u h}{u}+\frac{u^{2z-2}\phi^2}{r^{2z}_{+} h}-\frac{2(z+1)h}{u^2}-\frac{L^2m^2}{u^2}~\right] F(x,u;p)
%\nonumber \\
%  =\frac{1}{r^2_{+}} \left[ - \partial_{x}^2 + \left( p + B_{c_2} x \right)^2~\right] F(x,u;p) .
%\label{e5}
%\end{align}%
%Then we separate the $F$ as $F_n(x,u; p)=\rho_n(u)\gamma_n(x;p)/L$, where
%$\lambda_n$ is a constant. $\rho_n$ and $\gamma_n$ admit the following equations:
\begin{subequations}
\begin{align}
  \left( - \frac{\partial^2}{\partial X^2} + \frac{X^2}{4} \right) \gamma_n(x;p)& =\frac{\lambda_n}{2}\, \gamma_n(x;p) ,
\label{peom1} \\
  \left[h\partial_{u}^2 +(\partial_{u} h+\frac{(3-z)h}{u})\partial_{u}\right] \rho_{n}(u)
 & = \left(\frac{m^2 L^2}{u^2}+\frac{2(z+1)h}{u^2}-\frac{2\partial_u h}{u}-\frac{u^{2z-2}\phi^2}{r^{2z}_{+}h}+\frac{B_{c_2} \lambda_n}{r^2_{+}}\right)\rho_n(u) ,
\label{peom2}
\end{align}
\end{subequations}
where $X \equiv \sqrt{2 B_{c_2}} ( x + p/B_{c_2} )$.  The distribution of the order parameter on
the $(x-y)$ plane is given by the solution of the transverse equation
Eq. (\ref{peom1}). On the other hand, the radial equation Eq. (\ref{peom2}) determines superconducting phase transition.

\subsection{The upper critical magnetic field}
There is critical value $B_{c_2}$ above which Eq. (\ref{peom2}) only has vanishing solutions. As one lowers the magnetic field below $B_{c_2}$, we lead to a phase transition.
The maximum upper critical magnetic field is given by $n=0$ where  $\lambda_n=2n+1$ takes the minimum value. Thus, we can
express the equation of $\rho(u)$ as
\begin{eqnarray}
\rho''+\left[\frac{h'}{h}+\frac{3-z}{u}\right]\rho'+\left[\frac{2h'}{uh}+\frac{u^{2z-2}\phi^2}{r^{2z}_{+}h^2}
 -\frac{2(z+1)}{u^2}-\frac{m^2L^2}{u^2h}-\frac{B_{c2}}{r^2_{+}h}\right]\rho=0. \label{peom3}
\end{eqnarray}
Near boundary ($u \rightarrow 0$), $\rho$ in (\ref{peom3}) behaves like
\begin{equation}
 \rho(u)=J_{-}u^{\Delta_{-}}+J_{+}u^{\Delta_{+}}.\label{f4}.
\end{equation}
As before we let $J_{-}=0$ and set $J=J_{+}$ and $\Delta=\Delta_{+}$ in the following discussions.

Using the matching method just what we did in the last section, one can get from  Eq.(\ref{peom3})
\begin{eqnarray}
 & p B_{c_2}^2 + r^2_{+}\left\{\left[8(z+2)+2m^2L^2\right]p+4(z+2)q\right\}B_{c_2}+r^4_{+}\bigg\{\left[m^4L^4+10(z+2)m^2L^2\right. \nonumber \\
 &\left.+14(z+2)^2\right]p+\left[8(z+2)^2+4(z+2)m^2L^2\right]q+4(z+2)^2\Delta-p\alpha^2\bigg\}=0,
\end{eqnarray}
in which $p=\left[2u_m+(1-u_m)\Delta\right](1-u_m)$ and $q=\left[u_m+(1-u_m)\Delta\right]$.

In order to get the external critical magnetic field which is very close to the critical magnetic field we find the solution
\begin{equation}
 B_{c_2}=\frac{r^2_{+}}{2p}\left(\sqrt{\gamma+4p^2\alpha^2}-\beta\right),\label{BC2e}
\end{equation}
with
\begin{eqnarray}
 &\gamma=8(z+2)\left[(z+2)-m^2L^2\right]p^2+32(z+2)^2pq+16(z+2)^2q^2-16(z+2)^2\Delta
 p,
\end{eqnarray}
\begin{equation}
\beta=\left[8(z+2)+2m^2L^2\right]p+4(z+2)q.
\end{equation}

As $T=\frac{(z+2)r^z_{+}}{4\pi L^2}$, we can express the critical magnetic field $B_{c_2}$ as
\begin{eqnarray}\label{Bc}
 & B_{c_2} &=\left(\frac{4\pi T_c}{z+2}\right)^{\frac{2}{z}}\frac{1}{2p}\nonumber \\
 &&\left(\sqrt{(\beta^2-\gamma)\left[1+(z-1)(1-u_m)\right]^2 u^{2z-2}_m+\gamma\left(\frac{T}{T_c}\right)^{\frac{4}{z}}}-\beta\left(\frac{T}{T_c}\right)^{\frac{2}{z}}\right),
\end{eqnarray}
which has the same form as \cite{Pan}. %
%
%What interests us most is that we get the relation when we put the formula of $\beta$, $\gamma$, $\alpha$ and $p$,
%\begin{equation}
%\beta^2-\gamma=4\alpha^2p^2
%\end{equation}
%so
It is convenient to observe that there is a superconducting phase
transition when $B_{c_2}=0$ at $T=T_c$ where
\begin{equation}\label{beta}
\gamma=\beta^2.
\end{equation}
This is equivalent to
\begin{equation}\label{um}
u^{2z-2}_m \left[1+(z-1)(1-u_m)\right]^2=1,
\end{equation}
which is related to Lifshitz scaling but independent of the tensor
field mass. In order to ensure the condition $B_{c_2}=0$ at $T=T_c$,
we calculate the Eq.(\ref{beta}) and Eq.(\ref{um}) with the
requirement of $m^2L^2\geq -\frac{(2-z)^2+8(z+1)}{4}$. As a consequence, we get
\begin{eqnarray}
\left\{\begin{array}{lcr}
0<u_m<1, & \text{as} & z=1,\\
u_m=1,   & \text{as} & z\neq 1.
\end{array}\right.
\end{eqnarray}
What can be noted is that all of the $u_m$ is selected in the range $0<u_m<1$ with the situation $z=1$ so we can choose the matching point $u_m$ arbitrarily for this case. And the results also show that the allowable range of $u_m$ is restricted at the point $u_m=1$ when $z=\frac{3}{2}$. So we clearly find  that the range of the matching point $u_m$ depends on Lifshitz scaling $z$ and tensor field mass $m$. %and the range of $u_m$ becomes smaller as we amplify the value of $z$ for fixed $d$, the conclusion is also gained in \cite{Pan}.
\begin{figure}[htbp]
 \begin{minipage}{1\hsize}
\begin{center}
%\vspace*{10mm}
\includegraphics[scale=0.6]{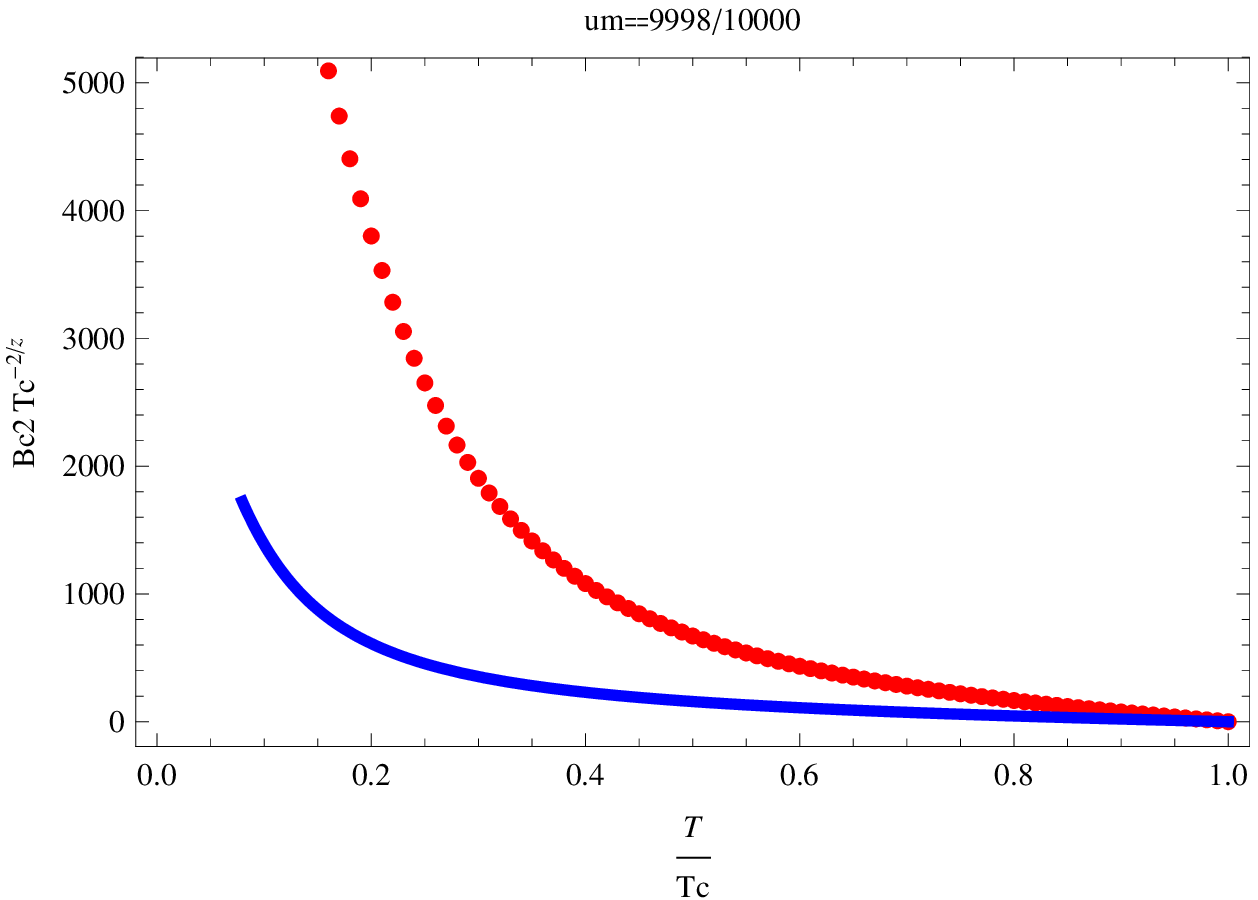}
\includegraphics[scale=0.6]{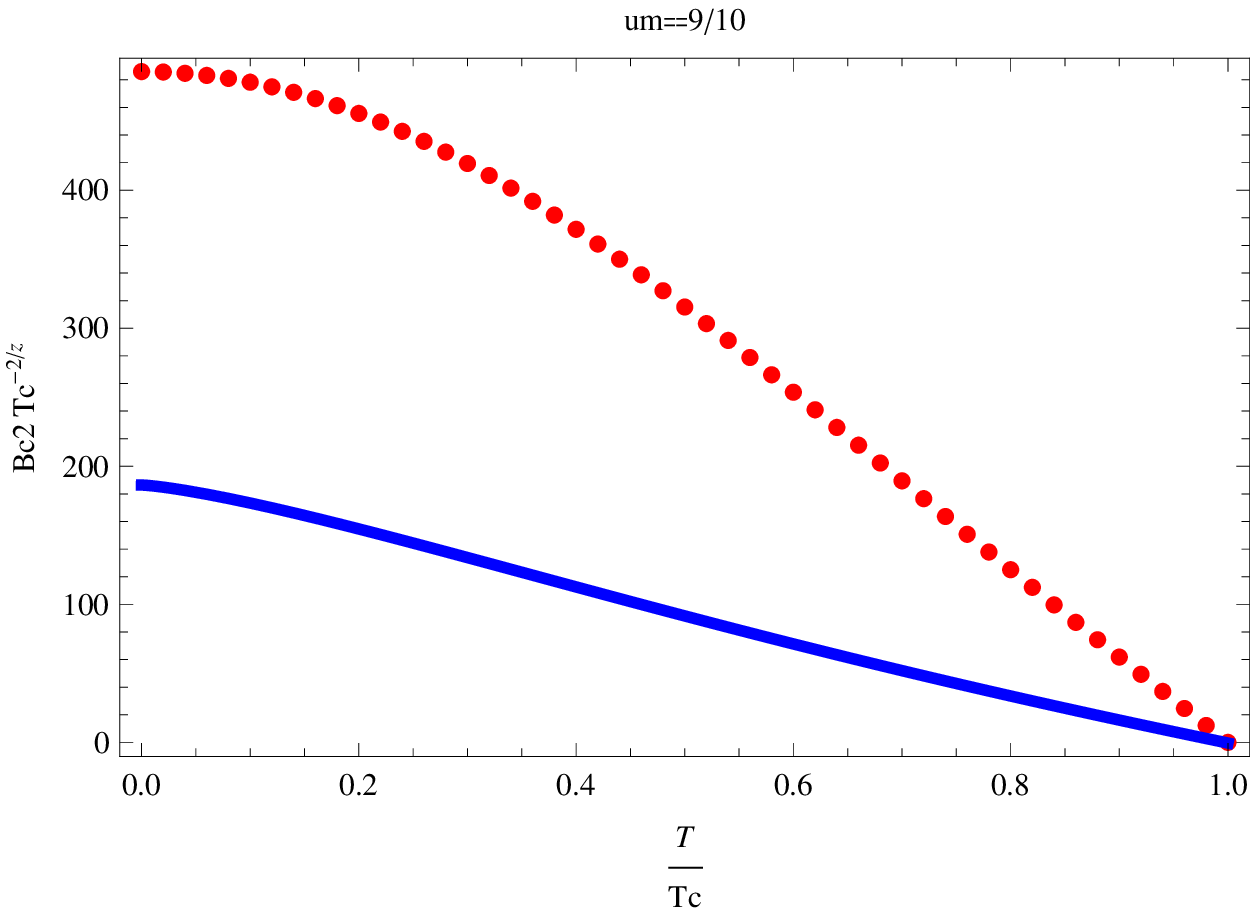}
\end{center}
\caption{ \label{Fig02} The critical magnetic field as a function of $T/T_c$ obtained by the analytic matching method. The left graph we set $u_m=9998/10000$, it shows that there is a breakdown of matching method. The right figure refers to $u_m=9/10$, which has more reasonable behavior. As before, in both graphs red point line denotes $z=1$ and blue bold line describes $z=3/2$.}
\end{minipage}
\end{figure}

It is subtle for $z\neq 1$ where the matching point $u_m=1$ leads to vanishing critical temperature $T_c$ as shown in (\ref{T_c}) and \eqref{alpha}, which implies a breakdown of the matching method.
%which implies that we cannot obtain the correct expression of the
%critical magnetic field $B_{c_2}$.
Our strategy is to matching the result by shifting the matching
point $u_m$ to a small value $\delta$ from $1$. However, $\delta$
cannot be arbitrarily small. As $u_m \rightarrow 1$, the value of
$p$ approaches to zero, which causes $B_{c_2}$ divergent. As an
example, we choose the Lifshitz scaling $z=3/2$ with $L=1$,
$m^2=-1/4$ and $r_+=1$, the left graph of Fig.\ref{Fig02} shows that
there exists a breakdown of matching method when
$u_m=\frac{9998}{10000}$.

Keep this in mind, we should choose $\delta$ is large enough so as
to keep $B_{c_2}$ finite. As another example, we set $u_m=9/10$
which can relax the breakdown when the matching point approaches the
allowable region. The right graph of Fig.\ref{Fig02} proves this
point and shows that the critical magnetic field $B_{c_2}$ decreases
as we amplify $z$ which is qualitatively in good agreement with the
work of \cite{Pan}. When $T \thicksim T_c$ we can have
$B_{c_2}\varpropto (1-T/T_c)$ for different Lifshitz scaling which
agrees well with the Ginzburg-Landau theory. And it is also noted
that the dynamical exponent $z$ cannot modify the relation. Thus,
for the case $1\leqslant z<d=2$, the Ginzburg-Landau theory still
holds in Lifshitz black hole.

\section{Vortex lattice of the $d$-wave superconductor.}
Based on our previous observations in the last sections, in this section we would like to construct the vortex solution of $d$-wave superconductor model, following the work \cite{Maeda2}. The main motivation is to see whether and how the dynamical exponent $z$ could impose its influences on the formation of the lattice.
\subsection{Leading order solution}
As a first step, we will consider the leading order ($\epsilon^{1/2}$) solution for the field $f(\textbf{x},u)$. The next order corrections will be discussed in the next subsection.  Our start point is  Eq. (\ref{peom1}), whose solution of that
satisfies the boundary conditions is the following
\begin{align}\label{dropsol}
\gamma_n(x;p) = e^{- X^2/4 } H_n(X),
\end{align}
where $\lambda_n = 2 n +1, (n=0,1,2,3 \cdots)$ are the corresponding eigenvalues. It was noted in \cite{Maeda2} that a vortex lattice solution of Eq. \eqref{peom1} can be constructed by linearly combining the lowest solution of \eqref{dropsol} through
\begin{equation}
 f_1(\textbf{x},u)
  = \frac{\rho_{0}(u)}{L^2}
   \sum_{\ell} c_{\ell}\, e^{i p_{\ell} y} \gamma_{0}(x; p_{\ell}),
\label{vsol2}
\end{equation}
where $\rho_0(u)$ is the solution of radial equation \eqref{peom3} and $\gamma_0(x;p)$ is the droplet solution which is given by
\begin{align}
  & \gamma_0(x;p)
  = \exp\left[ - \frac{1}{2 r_0^2}
    \left(x + p r_0^2 \right)^2 \right],
\label{vsol1}
\end{align}
with $r_0 \equiv 1/\sqrt{B_{c_2}}$.

It should be noted that the above solution (\ref{vsol1}) is very similar to the expression of the order
parameter in GL theory for the type II superconductor in the
presence of a magnetic field
\begin{equation}
\psi_L=\sum_{l} c_l e^{i p_l y}\textrm{ exp}[-\frac{x-x_l}{2 \xi^2}],\label{vsol3}
\end{equation}
where $\xi$ is the superconducting coherence length, $x_l=\frac{k \Phi_0}{2\pi B}$, and $\Phi_0$ is the flux quantum. As a result, Eqs. (\ref{vsol2}) and (\ref{vsol3}) gives
\begin{equation}
B_{c_2}\propto \frac{1}{\xi^2},
\end{equation}
which is the same as the result of the GL theory. In the previous section we have $B_{c_2}\propto(1-T/T_c)$ near $T_c$, which indicates that $\xi
\propto (1-T/T_c)^{-1/2}$ and is also the same as that of the GL theory.

Just like what has been made for the $s$-wave model in Ref. \cite{Maeda2}, one can construct the vortex lattice from droplet solutions by considering the following superposition:
\begin{subequations}
\begin{align}
  & f_1(\textbf{x},u)=\frac{\rho_0(u)}{L^2} \gamma_{L} (\textbf{x})
  = \frac{\rho_0(u)}{L^2}
  \sum_{\ell=-\infty}^{\infty} c_{\ell}\, e^{i p_{\ell} y} \gamma_0(x; p_{\ell}) ,
\label{g4} \\
  & c_{\ell} \equiv \exp\left( - i \frac{\pi a_2}{a_1^2} \ell^2 \right),
  \hspace{1.0truecm}
   p_{\ell} \equiv\frac{2 \pi \ell}{a_1 r_0},
\label{g5}
\end{align}
\label{g6}
\end{subequations}

\noindent for arbitrary parameters $a_1$ and $a_2$. The function $\gamma_L$ can be written as the elliptic theta function $\theta_3$ and has translation invariance (up to a phase)
in two directions ${\textbf{a}}=a_1 r_0\partial_y$ and
${\textbf{b}}=2\pi r_0/a_1\partial_x+a_2 r_0/a_1\partial_y$, and hence is called a vortex lattice. The
area of a unit cell for this vortex lattice is $2\pi r_0^2$, and the magnetic flux penetrating
the unit cell is given by $B_{c_2} \times (\text{Area}) = 2 \pi$, implying quantization of the magnetic flux penetrating a
vortex.

\begin{figure}[htbp]
 \begin{minipage}{1\hsize}
\begin{center}
%\vspace*{10mm}
\includegraphics*[scale=0.8] {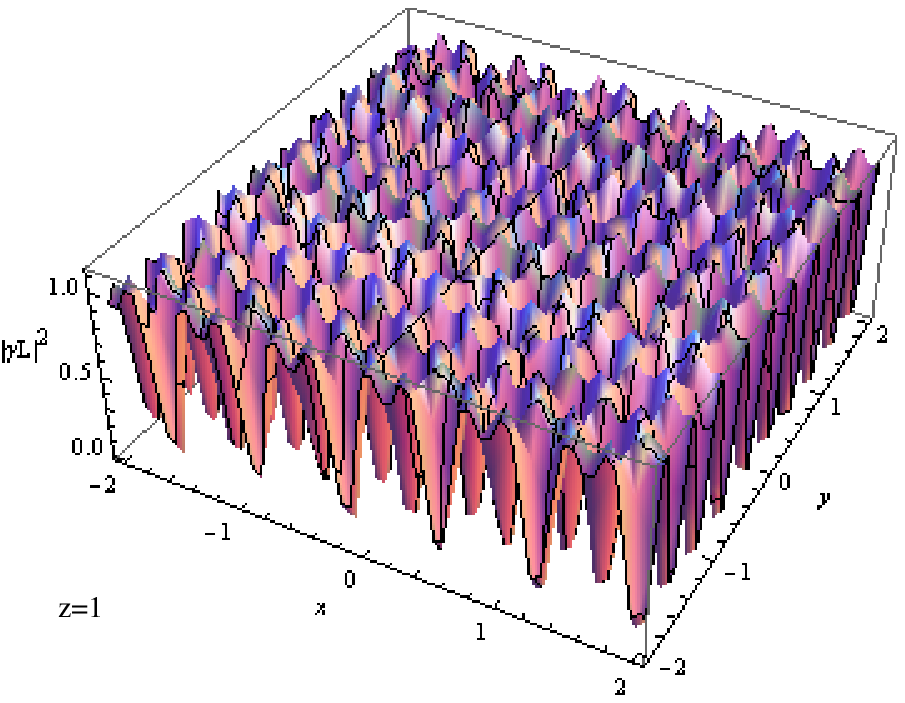}
\includegraphics*[scale=0.8] {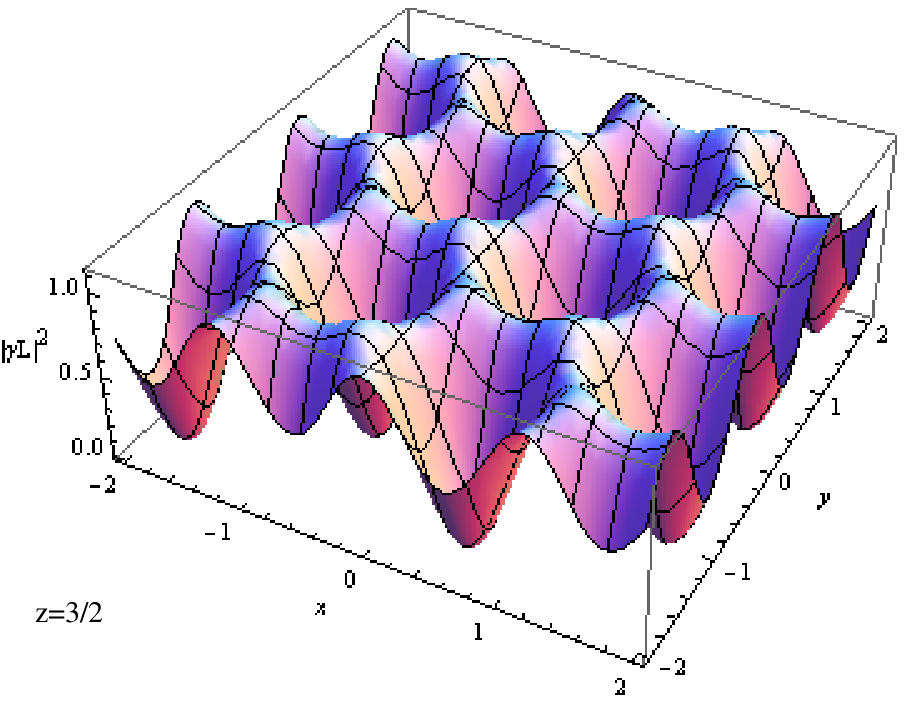}
\end{center}
\caption{\label{Fig03} The vortex lattice structure for the triangular lattice in the $(x,y)$ plane. The vertical line represents $\sigma=|\gamma_L|^2$, and vortex cores are located at $|\gamma_L|=0$. }
\end{minipage}
\end{figure}

\begin{figure}[htbp]
 \begin{minipage}{1\hsize}
\begin{center}
%\vspace*{10mm}
\includegraphics[scale=0.8]{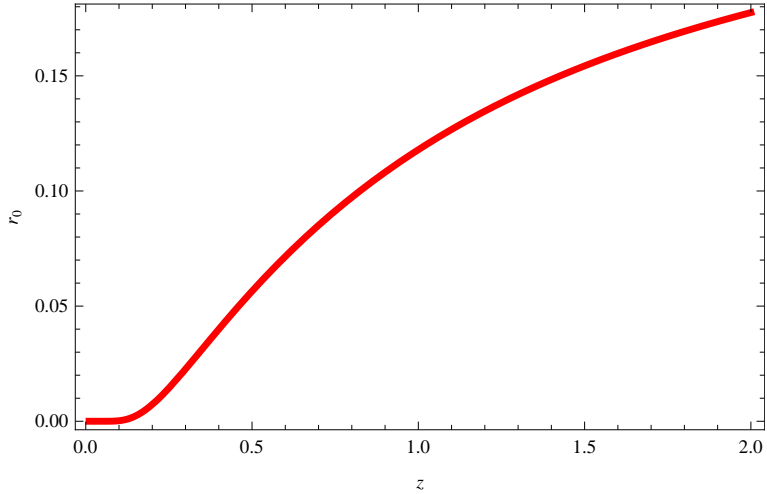}
\end{center}
\caption{ \label{Fig04} The characteristic scale $r_0$ of the unit cell plotted as a function of $z$.}
\end{minipage}
\end{figure}

Fig.\ref{Fig03} shows the configuration of $\sigma(\textbf{x})=|\gamma_{L}(\textbf{x})|^2$ in the $(x,y)$ plane for the rectangular lattice. In this plot we have chosen $u_m=9/10$, $m^2L^2=-1/4$, $r_+=1$ and $\rho=80$ for the Lifshitz scaling $z=1$ and $z=3/2$ respectively. These plots reveal a very interesting fact that the critical dynamical exponent $z$ cannot affect the shape of the lattice, instead it only changes the characteristic length $r_0$ (which is proportional to the coherence length $\xi$ of the superconductor) of the unit cell. Specifically, as shown in Fig. (\ref{Fig04}), the characteristic length $r_0$ (equivalently the coherence length $\xi$) increases with $z$ monotonously.
%ing that from the graphs the quantity of vortex lattices decreases when we amplify $z$ with the sizes becoming bigger although the height of $\sigma$ is not modified. It indicates that the dynamical exponent $z$ influences the vortex lattice solution but it can not modify the magnetic field in the holographic superconductors.

\subsection{Higher order corrections}
One of the most interesting result in the last subsection is that the dynamical exponent $z$, at least in the leading order, dose not influence the droplet solution and hence the form of the vortex lattice. There has the same result in holographic $s-$wave superconductor as shown in \cite{Lala}. However, though it is correct in the first order solution, whether it is still hold or not in higher order expansions is not clear. As a preliminary check, in this subsection we will pay our attention to the question to the next order, i.e., $\epsilon^{3/2}$ in $f(\textbf{x},u)$ expansions.  To do so, we expand the fields in the way like \eqref{expansion} and write down the equations of motion to the order $\epsilon^{3/2}$:
\begin{eqnarray}\label{f201}
& h\partial_{u}^2 f_2 + (\partial_{u} h +\frac{3-z}{u} h)\partial_{u} f_2 +\frac{1}{r^2_{+}}(\partial_{x}^2+\partial_{y}^2)f_2
  + \frac{2 i Ay^{(0)}}{r^2_{+}}\partial_{y} f_2 + \bigg\{\frac{2\partial_{u} h}{u} + \frac{u^{2z-2}(\phi^{(0)})^2}{r^{2z}_{+}h} -\frac{2(z+1)h}{u^2} \nonumber \\
&-\frac{{Ay^{(0)}}^{2}}{r^2_{+}}-\frac{m^2L^2}{u^2}\bigg\}f_2 + \frac{2 i Ay^{(1)}}{r^2_{+}}\partial_{y} f_1
  + (\frac{2u^{2z-2}\phi^{(0)}\phi^{(1)}}{r^{2z}_{+}h}-\frac{2Ay^{(0)}Ay^{(1)}}{r^2_{+}})f_1  =0,
\end{eqnarray}
where the zeroth order fields $Ay^{(0)}$ and $\phi^{(0)}$ have been obtained in the previous sections and the $\epsilon^{1/2}$ order solution $f_1$ is given by Eq. \eqref{g4}. In addition, there are $\epsilon$ order fields $Ay^{(1)}$ and $\phi^{(1)}$ which can be obtained by solving the following differential equations:
\begin{subequations}
\begin{eqnarray}\label{Ay101}
h\partial_{u}^2 Ay^{(1)}+(\partial_{u} h - \frac{z-1}{u}h)\partial_{u} Ay^{(1)}
+\frac{1}{r^2_{+}}\partial_{x}^2 Ay^{(1)}&=&\frac{4u^2}{r^4_{+}L^2}(p_{l}+Ay^{(0)})F_{1}^2,\\
\label{phi101}
h\partial_{u}^2 \phi^{(1)}+\frac{(z-1)h}{u}\partial_{u} \phi^{(1)}
+\frac{1}{r^2_{+}}(\partial_{x}^2+\partial_{y}^2) \phi^{(1)}&=&\frac{4u^2F_{1}^2\phi^{(0)}}{r^4_{+}L^2}
\end{eqnarray}
\end{subequations}
where $F_{1}\equiv \frac{\rho_0(u)}{L^2} \sum_{\ell} e^{ - \frac{1}{2r^2_{0}}(x+p_{\ell} r^2_{0})^2}$ and as before $p_{\ell}=\frac{2\pi \ell}{a_1 r_0}$.

The above Eqs. \eqref{Ay101} and(\ref{phi101}) can be solved analytically in the following way:

(i) We first note that the right sides of these two equations are independent of $y$, implying both $Ay^{(1)}$ and $\phi^{(1)}$ are $y$-independent.

(ii) Notice that the lattice has periodicity as $\frac{2\pi}{a_1 r_0}$ in $x$ direction, we therefore expand $A_y^{(1)}$ and $\phi^{(1)}$ as a Fourier series in $x$ coordinate\footnote{Theoretically one cannot determine the lattice shape due to the free parameters $a_1$ and $a_2$. For simplify, we would like to set $a_2=0$ in the following discussions. By doing so, we have chosen a rectangular lattice for the following consideration.
},
\begin{subequations}\label{fourier}
\begin{eqnarray}
A_y^{(1)}(x,u)&=&\frac{2a_1}{\sqrt{\pi}L^6}\sum_{k,\ell}\mathcal{D}(k,\ell) e^{\frac{ia_1 kx}{r_0}}e^{-\frac{k^2 a^{2}_{1} }{4}+i\pi k \ell - \frac{\pi^2 \ell^2}{a^{2}_{1}}}\widetilde A(k,\ell,u),\\
\phi^{(1)}(x,u)&=&\frac{2a_1}{\sqrt{\pi}L^6}\sum_{k,\ell} e^{\frac{ia_1 kx}{r_0}}e^{-\frac{k^2 a^{2}_{1} }{4}+i\pi k \ell - \frac{\pi^2 \ell^2}{a^{2}_{1}}}\widetilde \phi(k,\ell,u),
\end{eqnarray}
\end{subequations}
where $\mathcal{D}(k,\ell)\equiv (2\pi\ell+i k a_1^2)/(2a_1)$.

(iii) Using the Fourier series relations \eqref{A1} and \eqref{A2},
the set of equations \eqref{Ay101} $~$ (\ref{phi101}) reduce to
\begin{subequations}\label{Ay102}
\begin{eqnarray}
h \widetilde A''+ (h' - \frac{z-1}{u}h) \widetilde A'
- \frac{k^2 a^2_1}{r^2_{+}r_0^2} \widetilde A+\frac{ u^2\rho_0^2(u)}{r_0r^4_{+}}&=&0,\\
\label{phi102}
h\widetilde \phi''+\frac{(z-1)h}{u}\widetilde \phi'
-\frac{k^2 a^2_1}{r^2_{+}r_0^2} \widetilde \phi-\frac{ u^2 \rho_0^2(u)}{r^4_{+}}\phi^{(0)}&=&0,
\end{eqnarray}
\end{subequations}
where prime denotes differentiate w.r.t. $u$ and $\rho_0(u)$ again is the solution of \eqref{peom3}.
In general, it is not easy to analytically solve these differential equations. However, the asymptotic behaviour near the boundary $u \rightarrow 0$ is very straightforward
\begin{equation}\label{Ay102}
\widetilde A(u,k,\ell)= \frac{ J^2 u^{2\Delta+4}}{r_0r^4_{+}\left[(2\Delta+4)z-(2\Delta+4)^2\right]},
\end{equation}
\begin{equation}\label{phi102}
\widetilde \phi(u,k,\ell)=\frac{  J^2 u^{2\Delta+4}}{r^4_{+}(2\Delta+4)}\left(\frac{\mu}{2\Delta+2+z}-\frac{\rho u^{2-z}}{2\Delta+6-z}\right),
\end{equation}
which are independent of $k$ and $\ell$. Their behaviors far from the boundary can be obtained by solving Eqs. \eqref{Ay102} numerically.  One thing should be mention is that the coefficients in the Fourier series \eqref{fourier} are exponentially suppressed as a function of $k^2$ and $\ell^2$. This point and the fact that $u<1$ lead to a consequence that one can neglect the terms proportional to $k^2$ in Eqs. \eqref{Ay102}, then they become
\begin{subequations}\label{Ay103}
\begin{eqnarray}
h \widetilde A''(u)+ (h' - \frac{z-1}{u}h) \widetilde A'(u)+\frac{u^2\rho_0^2(u)}{r_0 r^4_{+}}&\simeq &0,\\
h\widetilde \phi''(u)+\frac{(z-1)h}{u}\widetilde \phi'(u)-\frac{ u^2 \rho_0^2(u)}{r^4_{+}}\phi^{(0)}&\simeq&0.
\end{eqnarray}
\end{subequations}
The above two equations can be solved analytically. Let us denote them by $\hat{A}(u)$ and $\hat{\phi}(u)$ (their exact expressions are irrelevant to our present discussion). After insert them into \eqref{fourier}, one obtains the full functions of $A_y^{(1)}(x,u)$ and $\phi^{(1)}(x,u)$
\begin{subequations}\label{Ay104}
\begin{eqnarray}
A_y^{(1)}(x,u)&=&\frac{4\hat{A}(u)}{r_0L^6}\sum_{k,\ell} \left(x+\frac{2\pi k r_0}{a_1}\right)e^{-\frac{1}{2r^2_0}\left(x+\frac{2\pi k r_0}{a_1}\right)^2-\frac{1}{2r^2_0}\left(x+\frac{2\pi (k+\ell) r_0}{a_1}\right)^2},\\
\phi^{(1)}(x,u)&=&\frac{4\hat{\phi}(u)}{L^6}\sum_{k,\ell} e^{-\frac{1}{2r^2_0}\left(x+\frac{2\pi k r_0}{a_1}\right)^2-\frac{1}{2r^2_0}\left(x+\frac{2\pi (k+\ell) r_0}{a_1}\right)^2},
\end{eqnarray}
\end{subequations}
where we have used the relations \eqref{A1} and \eqref{A2}.

The function $f_2$ can be solved in the same way. We first expand $f_2$ as a double Fourier series in $x$ and $y$
\begin{eqnarray}
f_2(x,y,u)&=&\frac1{L^8}\sum_{k,\ell,w}\mathfrak{F}(k,\ell,w)e^{\frac{2\pi iky}{a_1r_0}} e^{\frac{ia_1 kx}{r_0}}\tilde{f}_2(k,\ell,w,u).
\end{eqnarray}
Then we put Eq. \eqref{g4} and the series \eqref{fourier} into Eq.(\ref{f201}), making use of the relations of the infinite sum of Gaussians and the infinite sum of exponential in the Appendix, neglecting the terms proportional to $k$ and $\ell$, then we lead to the following differential equation for $f_2$ in the radial coordinate
\begin{eqnarray}\label{f2}
  & h \tilde{f}_2'' + ( h' +\frac{3-z}{u} h)\tilde{f}_2'
  +  \bigg\{\frac{2h'}{u} + \frac{u^{2z-2}(\phi^{(0)})^2}{r^{2z}_{+}h} -\frac{2(z+1)h}{u^2} -\frac{m^2L^2}{u^2}\bigg\}\tilde{f}_2+\frac{8 u^{2z-2}\phi^{(0)}\hat{\phi}(u)\rho_0(u)}{r^{2z}_{+} h}\simeq0,
  \end{eqnarray}
If we use the relation \eqref{A3} again, we can obtain an approximate solution for $f_2$
\begin{eqnarray}
f_2(x,y,u)&\simeq&\frac{\hat{f}_2(u)}{L^8}\sum_{k,\ell,w}\mathfrak{F}(k,\ell,w)e^{\frac{2\pi iky}{a_1r_0}}  e^{-\frac{1}{2r^2_0}\left(x+\frac{2\pi k r_0}{a_1}\right)^2-\frac{1}{2r^2_0}\left(x+\frac{2\pi (k+\ell) r_0}{a_1}\right)^2-\frac{1}{2r^2_0}\left(x+\frac{2\pi w r_0}{a_1}\right)^2},
\end{eqnarray}
where $\hat{f}_2(u)$ is a solution of \eqref{f2}. It is clear that the Lifshitz scaling does NOT affect the shape of the vortex just like the $\epsilon^{1/2}$ order.

\section{Conclusions and discussions}
In this work, we have used the matching method to investigate the $d$-wave holographic superconductors with Lifshitz scaling in the presence of external magnetic field. Based on purely analytic methods, the vortex lattice solutions of the model have also been obtained with different Lifshitz scaling. This implies that holographic $d$-wave superconductor, regardless of the anisotropy between space and time, is indeed a type II one, which is the same as the conventional $d$-wave superconductors in the GL theory. Our results also indicates that the dynamical exponent $z$ does NOT influence the shape of vortex lattice even after the higher order corrections are taken into consideration. However, it has effects to the structure of the vortex lattices through the characteristic length $r_0$. Also, close comparisons between our results and those of the GL theory reveal the fact that the upper critical magnetic field $B_{c_2}$ is inversely proportional to the square of the superconducting coherence length $\xi$. %but it can not modify the magnetic field in the holographic superconductors.%effectiveness of this analytic method.

Working in the probe limit, we obtained analytic expressions for the order parameter, the critical temperature and the upper critical magnetic field. The analytic calculation is useful for gaining insight into the strong interacting system. It is noted that the critical temperature decreases with the increase of the dynamical exponent $z$ showing that Lifshitz scaling makes the condensation harder to occur. In the section of the critical magnetic field, we also observed the behavior satisfying the relation given in the GL theory. The result shows that the dynamical exponent $z$ does have effects on the upper critical magnetic field based on the facts that a larger $z$ results in a smaller upper critical magnetic field. %Then we focus on the vortex lattices solution with the Lifshitz scaling in the holographic superconductor, the result satisfies the conclusion which we have discussed in the section 3 that the dynamical exponent $z$ does have effects on the upper critical magnetic field.

Although we have performed detailed analysis on some issues of holographic Lifshitz $d$-wave superconductor in the presence of external magnetic field, it would be many more interesting outcomes that deserve further investigations. Some of these are as follows: (i) It would be natural to generalize our discussions to Fermion field and to see how the dynamic exponent $z$ influence its condensation and vortex lattice solutions (There is some related work such as Ref. \cite{fb}, where the authors studied the fermionic wavefunctions for the relativistic $d$-wave superconducting background and found the formation of Fermi arcs. ). (ii) It would be possible to find analogous vortex lattice solutions for the hyperscaling violation holographic models. (iii) One significant difference between the conventional superconductors in the GL theory and the holographic superconductor hide in the free energy and the $R$-current. It is interesting to obtain these two quantities for our model and study the effects of anisotropy on them. (iv) It is also possible to consider the holographic superconductor model in the framework of modified gravity, such as the Ho\v{r}ava-Lifshitz (HL) gravity\cite{HL} proposed recently by Ho\v{r}ava. Indeed, it was found that HL gravity is a minimal holographic dual for the fiedl with Lifshtiz scaling\cite{HL2}. Our recent works \cite{shu1,shu2} further revealed this point and found that various Lifshtiz spacetimes are possible even without matter fields. It is of particular interests to see how to construct the holographic superconductor models in this framework.

\section*{\bf Acknowledgements}

We would like to thank X.-H. Ge for useful discussions. This work was supported in part by the National Natural Science Foundation of China (under Grant Nos. 11465012 and 11005165), the Natural Science Foundation of Jiangxi Province (under Grant No. 20142BAB202007) and the 555 talent project of Jiangxi Province.

\appendix

%%%%%%%%%%%%%%%%%%%%%%%%%%%%%%%%%%%%%%%%%%%%%%%%%%%%%%%%%%%%%%%%%
%%%%%%%%%%%%%%%%%%%%%%%%%%%%%%%%%%%%%%%%%%%%%%%%%%%%%%%%%%%%%%%%%

\section{Useful relations} \label{appendixA}

%%%%%%%%%%%%%%%%%%%%%%%%%%%%%%%%%%%%%%%%%%%%%%%%%%%%%%%%%%%%%%%%%
%%%%%%%%%%%%%%%%%%%%%%%%%%%%%%%%%%%%%%%%%%%%%%%%%%%%%%%%%%%%%%%%%
By making use of the properties of the infinite sum of Gaussians and the infinite sum of exponential as shown in \cite{1303.4390}, we
\begin{eqnarray}\label{A1}
&&\sum_{k} e^{-\frac{1}{2r^2_0}\left(x+\frac{2\pi k r_0}{a_1}\right)^2-\frac{1}{2r^2_0}\left(x+\frac{2\pi (k+\ell) r_0}{a_1}\right)^2}
=\sum_{k} e^{\frac{ia_1 kx}{r_0}}\mathfrak{E}(k,\ell)\\
\label{A2}
&&\sum_{k} \left(x+\frac{2\pi k r_0}{a_1}\right)e^{-\frac{1}{2r^2_0}\left(x+\frac{2\pi k r_0}{a_1}\right)^2-\frac{1}{2r^2_0}\left(x+\frac{2\pi (k+\ell) r_0}{a_1}\right)^2}
=-r_0\sum_{k} \left(\frac{\pi \ell}{a_1}+i\frac{ka_1}{2}\right)e^{\frac{ia_1 k x}{r_0}}\mathfrak{E}(k,\ell),\nonumber\\ \ \\
&&\sum_{k} e^{-\frac{1}{2r^2_0}\left(x+\frac{2\pi k r_0}{a_1}\right)^2-\frac{1}{2r^2_0}\left(x+\frac{2\pi (k+\ell) r_0}{a_1}\right)^2-\frac{1}{2r^2_0}\left(x+\frac{2\pi w r_0}{a_1}\right)^2}
=\sum_{k} e^{\frac{ia_1 kx}{r_0}}\mathfrak{F}(k,\ell,w)\label{A3}\\
&&\sum_{k}(x+\frac{2\pi kr_0}{a_1})x e^{-\frac{1}{2r^2_0}\left(x+\frac{2\pi k r_0}{a_1}\right)^2-\frac{1}{2r^2_0}\left(x+\frac{2\pi (k+\ell) r_0}{a_1}\right)^2-\frac{1}{2r^2_0}\left(x+\frac{2\pi w r_0}{a_1}\right)^2}\nonumber\\
&&~~~~~~~~~~~~~~~~~~~~~~~~=\sum_{k} \left(\frac{r_0^2}{3}+\mathfrak{N}^2-\mathfrak{M}\mathfrak{N}\right)e^{\frac{ia_1 kx}{r_0}}\mathfrak{F}(k,\ell,w),
\end{eqnarray}
where
\begin{eqnarray}
\mathfrak{E}(k,\ell)&=&\frac{a_1}{2\sqrt{\pi}}e^{-\frac{k^2 a^{2}_{1} }{4}+i\pi k \ell - \frac{\pi^2 \ell^2}{a^{2}_{1}}},\\
\mathfrak{F}(k,\ell,w)&=&\frac{a_1}{\sqrt{6\pi}} \exp{\left[\frac{4\pi^2}{3a_1^2}(2kw+w\ell-k\ell-k^2-\ell^2-w^2)+\frac{2\pi i (2k+\ell+w)k}{3}-\frac{k^2a_1^2}{6}\right]},\nonumber\\ \ \\
\mathfrak{M}&=&\frac{2\pi kr_0}{a_1},\\
\mathfrak{N}&=&\frac{2\pi (2k+\ell+w)r_0}{3a_1}+\frac{ika_1r_0}{3}.
\end{eqnarray}

%\section{the solution of the second order function} \label{appendixB}
%Substituting the Eq.(\ref{Ay104}) into Eq.(\ref{f201}), we expect that $f_2$ has the same form in $y$ direction as that of $f_1$. Therefore, we  can expand $f_2$ as a Fourier series in $x$ and $y$ coordinates in the following way:
%\begin{eqnarray}
%f_2(x,y,u)&=&\frac1{L^5}\sum_{k,\ell,w}\mathfrak{F}(k,\ell,w)e^{\frac{2\pi iky}{a_1r_0}} e^{\frac{ia_1 kx}{r_0}}\tilde{f}_2(k,\ell,w,u).
%\end{eqnarray}
%Eq. \eqref{f201} now becomes
%\begin{eqnarray}
%  & h \tilde{f}_2'' + ( h' +\frac{3-z}{u} h)\tilde{f}_2'
%  +  \bigg\{\frac{2h'}{u} + \frac{u^{2z-2}(\phi^{(0)})^2}{r^{2z}_{+}h} -\frac{k^2a_1^2}{r_0^2r_+^2}-\frac{4\pi^2 k^2}{a_1^2 r^2_0 r^2_{+}}-\frac{4\pi k x}{a_1 r^3_0 r^2_{+}}  \nonumber \\ £¦&-\frac{2(z+1)h}{u^2} - \frac{x^{2}}{r^4_0 r^2_{+}}-\frac{m^2L^2}{u^2}\bigg\}\tilde{f}_2+Z_1+Z_2+Z_3=0,
%  \end{eqnarray}
%  where
%\begin{eqnarray}
%  Z_1&=& \frac{16\pi k\hat{A}(u)\rho_0(u)}{a_1r_0^2r_+^2}(\mathfrak{N}-\mathfrak{M}),\\
%  Z_2&=& \frac{8 u^{2z-2}\phi^{(0)}\hat{\phi}(u)\rho_0(u)}{r^{2z}_{+} h},\\
%  Z_3&=&-\frac{8 \hat{A}(u)\rho_0(u)}{3r_0^3r^{2}_{+} }(r_0^2+3\mathfrak{N}^2-3\mathfrak{M}\mathfrak{N}).
%  \end{eqnarray}

\end{document}